\documentclass[prd,aps,a4paper,twocolumn,nofootinbib,showpacs]{revtex4-1}

\usepackage{graphicx,psfrag}
\usepackage{mathrsfs}
\usepackage{amsmath,amsfonts,amssymb}
\usepackage{multirow}
\usepackage{comment}
\usepackage{bm}
\usepackage{graphicx}
\usepackage{psfrag}
\usepackage{epstopdf}
\usepackage{color}
\usepackage{mathrsfs}
\usepackage{comment}
\usepackage{ulem}
\usepackage{hyperref}
\usepackage{diagbox} % for up-to date system 
\usepackage{amsmath,amssymb,amsthm,wasysym}
\usepackage{capt-of}
\usepackage[utf8]{inputenc}

\def\ha{{\hat{a}}}

\def\ha{{\hat{a}}}
\def\ph{\phi}

\newcommand{\be}{\begin{equation}}
\newcommand{\ee}{\end{equation}}

\newcommand{\RN}[1]{%
  \textup{\uppercase\expandafter{\romannumeral#1}}%
}

\definecolor{gray}{rgb}{0.5,0.5,0.5}
\definecolor{cyan}{rgb}{0,0.9,0.9}
\definecolor{orange}{rgb}{0.9,0.5,0}
\definecolor{magenta}{rgb}{1,0,1}
\definecolor{purple}{rgb}{0.8,0.4,0.8}
\definecolor{darkgreen}{rgb}{0,.6,0}
\definecolor{turquoise}{rgb}{0.25,0.88,0.82}

\newcommand{\RM}[1]{\MakeUppercase{\romannumeral #1{}}}

\begin{document}

\interfootnotelinepenalty=10000
\raggedbottom

\title{Spinning test-body orbiting around a Kerr black hole: \\
       circular dynamics and gravitational-wave fluxes }

\author{Georgios Lukes-Gerakopoulos${}^1$, Enno Harms${}^2$, Sebastiano Bernuzzi${}^3$, Alessandro Nagar${}^4$}
\affiliation{${}^1$ Astronomical Institute of the Academy of Sciences of the Czech Republic,
Bo\v{c}n\'{i} II 1401/1a, CZ-141 31 Prague, Czech Republic}
\affiliation{${}^2$Theoretical Physics Institute, University of Jena, 07743 Jena, Germany}
\affiliation{${}^3$DiFeST, University of Parma, and INFN, 43124, Parma, Italy}
\affiliation{${}^4$Institut des Hautes Etudes Scientifiques, 91440 Bures-sur-Yvette, France}

\begin{abstract}

In a recent work, [Phys.~Rev.~D. 94,~104010 (2016)], hereafter Paper~\RM{1},
we have numerically studied different prescriptions for the dynamics
of a spinning particle in circular motion around a Schwarzschild black hole.
In the present work, we continue this line of investigation to the rotating Kerr black hole.
We consider the Mathisson-Papapetrou formalism 
under three different spin-supplementary-conditions (SSC), the Tulczyjew SSC, the Pirani SSC
and the Ohashi-Kyrian-Semerak SSC, and analyze the different circular dynamics
in terms of the ISCO shifts and the frequency parameter ${x \equiv (M \Omega)^{2/3}}$,
where $\Omega$ is the orbital frequency and $M$ is the Kerr black hole mass.
Then, we solve numerically the inhomogeneous $(2+1)D$ Teukolsky equation to
contrast the asymptotic gravitational wave fluxes for the three cases. 
Our central observation made in Paper~\RM{1} for the Schwarzschild limit
is found to hold true for the Kerr background:
the three SSCs reduce to the same circular dynamics and the same radiation fluxes
for small frequency parameters but differences arise as $x$ grows close to the ISCO.
For a positive Kerr parameter $a=0.9$ the energy fluxes 
mutually agree with each other within a $0.2\%$ uncertainty up to $x<0.14$, 
while for $a=-0.9$ this level of agreement is preserved up to $x<0.1$.
For large frequencies ($x \gtrsim 0.1$), however, the spin coupling of the Kerr
black hole and the spinning body results in significant differences of the
circular orbit parameters and the fluxes, especially for the $a=-0.9$ case.
Instead, in the study of ISCO the negative Kerr parameter $a=-0.9$ results
in less discrepancies in comparison with the positive Kerr parameter $a=0.9$.
As a side result, we mention that, apart from the Tulczyew SSC,
ISCOs could not be found over the full range of spins: 
For $a=0.9$, for the Ohashi-Kyrian-Semerak SSC ISCOs could be found only for $\sigma<0.25$,
while for the Pirani SSC ISCOs could be found only for $-0.68<\sigma <0.64$.
For $a=-0.9$, for the Ohashi-Kyrian-Semerak SSC ISCOs could be found for $\sigma <0.721$.

\end{abstract}


\maketitle

\section{Introduction}
\label{sec:intro}

 The Mathisson-Papapetrou equations (MP)~\cite{Mathisson:1937zz,Papapetrou:1951pa}
 in the pole-dipole approximation describe the motion of a spinning test-body
 in a curved spacetime, once the centroid of the body is decided.
 A centroid is a single reference point inside the body, with respect to which 
 the spin is measured, and it is fixed by applying a spin
 supplementary condition (SSC). The worldline of the centroid represents the
 worldline of the extended test-body. From this point of view, the dynamics of
 the extended test-body is reduced to the dynamics of a point, and, therefore,
 the test-body is often called a test-particle.   
 
 There are different SSCs (see, e.g., \cite{Semerak:2015dza} for a recent review)
 and each SSC defines a different centroid for the body. Thus, one test-body can
 have different test particle descriptions.
 One would expect that the different descriptions should be equivalent.
 Indeed, the radius in which  
 different centroids have to lie in order to describe the same physical body 
 is equal to $S/\mu$ \cite{moller1949dynamique}, where $S$ is the measure of the
 spin and $\mu$ is the mass of the spinning body. This radius is known as the
 M\"{o}ler radius. Moreover, in~\cite{Kyrian:2007zz} it was shown how
 to describe the same body with two different SSCs. Namely, this is done by     
 shifting from the centroid of one SSC to the centroid defined by another SSC,
 and by performing certain transformations of the spin.
 Thus, if after such a shift the new centroid 
 lies inside M\"{o}ler radius, then the two centroids should describe the same
 body. Having these facts in mind, one would expect that a SSC
 just expresses a gauge freedom. While this intuition holds true for flat space,
 the issue gets more complicated on a curved background. Namely, it
 has been shown that worldlines that start as equivalent descriptions of the same
 body will diverge in such a way that after a while
 they cannot describe the same body anymore~\cite{Kyrian:2007zz}. 
 This implies that in the pole-dipole
 approximation a SSC is not exactly a gauge freedom, or equivalently, it implies that 
 the pole-dipole approximation fails in describing the dynamics of an extended body.
 In fact, if one takes all the multipoles of the body into account,
 then a SSC should be considered just a gauge~\cite{Costa:2015}.
 However, when truncating the multipole expansion at a certain order (in our case dipole), 
 the relations between the different SSCs can become more complex than just a
 gauge transformation, which motivates our study of their implications on the dynamics
 and gravitational wave emission.
 
 In Paper~\RM{1}, we have studied three different SSCs by analyzing their effects
 on the dynamics of a spinning particle in circular equatorial orbits (CEOs)
 around a Schwarzschild black hole.  In particular, we have compared quantities
 that should be invariant under gauge transformations, like e.g.~the ISCO, the 
 orbital frequency and the gravitational wave fluxes emitted to infinity.
 In the present work we expand our investigations to the Kerr background and 
 again examine CEOs as obtained with three different SSCs, i.e.~we consider the
 \begin{enumerate}
 \item[i)]~MP with the Tulczyew (T) SSC \cite{tulczyjew1959motion}, 
 \item[ii)]~MP with the Pirani (P) SSC \cite{Pirani:1956tn},
 \item[iii)]~MP with the Ohashi-Kyrian-Semerak (OKS) SSC \cite{Ohashi:2003,Kyrian:2007zz}.
 \end{enumerate}
 As in Paper~\RM{1}, we do not claim to consider the same extended test-body,
 when comparing the prescriptions i),~ii) and iii)
 because we do \textit{not} follow the mentioned procedure of centroid-shifting
 and spin-transforming explained in~\cite{Kyrian:2007zz}. Instead, for each SSC 
 we independently solve the equations of motion such that circular orbits
 are obtained and subsequently compare the dynamics and gravitational waves
 over some parameter like the particle spin or the orbital frequency.
 One of the reasons for this choice is that, given a circular orbit within one SSC,
 if one shifts the worldline in order to align with the centroid of another
 SSC, the obtained new worldline is not in general a CEO anymore.
 For example, if for the T SSC the worldline of the centroid is a CEO, then,
 after shifting the wordline to the P SSC centroid,
 the new centroid will most probably follow a helical motion~\cite{Costa:?}
 superposed on the averaged circular motion.

 The rest of the article is organized as follows. The theoretical foundations of
 the MP formalism and the procedures to find circular orbits are kept at a 
 minimal extent here since the respective discussions in Paper~\RM{1} were
 already presented in a general form valid for the Kerr spacetime. Thus, 
 Sec.~\ref{sec:MP_dynamics} only provides the most essential elements of the MP
 dynamics, in order to allow a more convenient reading, and presents the
 numerical findings for CEOs. The results for the ISCO shifts are discussed in
 Sec.~\ref{sec:ISCO}. The asymptotic gravitational wave fluxes are analyzed in 
 Sec.~\ref{sec:GWFlux}. Finally, Sec.~\ref{sec:Concl} summarizes the main 
 findings of this work.

\paragraph*{Units and notation:} 
 We keep all the conventions used in Paper~\RM{1}. Here we just briefly 
 mention the most elementary ones. Geometric units are employed throughout the
 work, i.e. ${G=c=1}$. The Riemann tensor is defined as
 ${{R^\alpha}_{\beta\gamma\delta}\equiv
 \Gamma^\alpha_{\gamma \lambda} \Gamma^\lambda_{\delta \beta}
 - \partial_\delta \Gamma^\alpha_{\gamma\beta}
 - \Gamma^\alpha_{\delta\lambda} \Gamma^{\lambda}_{\gamma\beta}
 + \partial_\gamma \Gamma^{\alpha}_{\delta \beta}}$,
 where the Christoffel symbols $\Gamma^\alpha_{\beta\gamma}$ are computed
 from the metric $g_{\alpha\beta}$ with signature $(-,+,+,+)$. Greek letters 
 denote the indices corresponding to spacetime (running from 0 to 3).
 $\epsilon_{\mu\nu\rho\sigma}=\sqrt{-g} \tilde{\epsilon}_{\mu\nu\rho\sigma} $
 denotes the Levi-Civita tensor with the Levi-Civita symbol set to be
 $\tilde{\epsilon}_{0123}=1$; $g$ is the determinant  of the metric tensor.
 In practice, we work numerically always with dimensionless quantities.
 Namely, by setting the black hole mass $M=1$, some dimensionful/dimensionless
 quantities become equivalent, e.g.~the Kerr spin parameter $a=\pm|\vec{S}_1|/M$,
 where $\vec{S}_1$ is the spin angular momentum, and its dimensionless version 
 $\ha=a/M$, as well as the radius $\hat{r}=r/M$. Furthermore, setting the
 test-body mass $\mu=1$ (or $\textsf{m}=1$ for the P SSC) the spin parameter
   \begin{align} \label{eq:dimless_quantities}
     \sigma=S/(\mu M)  
    \end{align}
 can be used interchangably with $S$.

\section{Mathisson-Papapetrou dynamics}
\label{sec:MP_dynamics}

 This section briefly reviews the MP equations and the characteristic features
 of CEOs.
 
 \subsection{Equations of motion} \label{sec:MP_EOM_SSC}
 
 The MP equations describe the evolution of a spinning particle's four momentum 
 and its spin-tensor. In their revised form~\cite{dixon1974dynamics}, they read
 \begin{subequations}
  \label{eq:MP}
  \begin{align}
   \frac{D~p^{\mu}}{d\lambda}&=-\frac{1}{2}~{R^{\mu}}_{\nu\kappa\lambda}v^{\nu}S^{\kappa\lambda}
    \quad , \label{eq:MPp}\\
    \frac{D~S^{\mu\nu}}{d\lambda}&=p^{\mu}~v^{\nu}-v^{\mu}~p^{\nu} \quad ,\label{eq:MPS}
  \end{align}
  \end{subequations}
  where ${D}/{d\lambda}$ denotes the covariant derivative along the 
  four-velocity $v^{\nu}$, since we choose $\lambda$ to be the proper time,
  i.e. $v_\nu v^\nu=-1$. By contracting Eq.~\eqref{eq:MPS} with the four-velocity,
  one gets
  \be \label{eq:hidmom}
   p^\mu=\textsf{m} v^\mu-v_\nu \frac{D~S^{\mu\nu}}{d\lambda} \quad ,
  \ee
  which shows that the four-momentum $p^{\nu}$ is in general not parallel to $v^{\nu}$,
  since the \textit{hidden} momentum
  \be
   p^\mu_{\rm{hidden}} \equiv-v_\nu \frac{D~S^{\mu\nu}}{d\lambda} \quad 
  \ee
  is in general non-zero. If the hidden momentum was zero, then the spin tensor
  $S^{\mu\nu}$ would be just parallel transported along the worldline, i.e. 
  \be
   \frac{D~S^{\mu\nu}}{d\lambda}=0 \quad .
  \ee
  The scalar $\textsf{m}\equiv -v_\nu p^\nu$ defines the mass with respect to the
  four-velocity, while the scalar $\mu \equiv \sqrt{-p_\nu p^\nu}$ defines
  the mass with respect to the four-momentum. While we have these two ways to define 
  a mass for the particle, the spin's measure is defined uniquely by
  \be \label{eq:spinMagnitude}
   S^2=\frac{1}{2}S^{\mu\nu}S_{\mu\nu} \quad .
  \ee
  
  Until this point we have not defined the centroid, i.e. the center of the mass,
  whose evolution in time forms the body's worldline. To define a centroid, we
  choose an observer represented by a future-pointing time-like vector $V^\mu$,
  for which it holds that
  \be \label{eq:SSC}
    V_\mu S^{\mu\nu}=0 \quad .
  \ee
  Condition~\eqref{eq:SSC} is the general form of a SSC. Moreover, without loss
  of generality we can choose $V_\mu$ to be the four-velocity of some time-like
  observer such that
  \be \label{eq:normV}
   V^\mu V_\mu=-1 \ \quad .
  \ee
  In particular, for the P SSC $V^\nu=v^\mu$, for the T SSC $V^\mu=p^\mu / \mu$, while
  for the OKS SSC $V^\mu$ is chosen such that $p^\mu_{\rm{hidden}}=0$.

  Once a SSC is imposed, it is possible to introduce a spin four-vector 
  \begin{align} \label{eq:SpinVect}
    S_\mu = -\frac{1}{2} \epsilon_{\mu\nu\rho\sigma}
          \, V^\nu \, S^{\rho\sigma} \quad,
  \end{align}
  whose inversion reads
  \begin{align}    \label{eq:T4VSin}
    S^{\rho\sigma}=-\epsilon^{\rho\sigma\gamma\delta} S_{\gamma}
    V_\delta \quad .
   \end{align}
   
   For a stationary and axisymmetric spacetime with a reflection symmetry along
   the equatorial plane (SAR spacetime), like the Kerr spacetime, we have two
   conserved quantities 
   \begin{align}
    E = -p_t+\frac12g_{t\mu,\nu}S^{\mu\nu} \label{eq:EnCons} \quad,\\
    J_z = p_\phi-\frac12g_{\phi\mu,\nu}S^{\mu\nu} \label{eq:JzCons} \quad,
   \end{align}
   where $t$ is the coordinate time, and $\phi$ the azimuthal angle of a 
   coordinate system, in practice Boyer-Lindquist coordinates, adapted to the
   symmetries. $E$ represents the energy and $J_z$ represents
   the component of the total angular momentum along the symmetry axis $z$ of
   the spacetime. The above two integrals are independent of the SSC, whereas
   the masses $\textsf{m}$,~$\mu$ and the measure of the spin $S$ in general
   depend on the SSC. For the three SSCs considered in this work the spin measure
   $S$ is a conserved quantity, while the mass $\textsf{m}$ is not a constant of
   motion only for the T SSC and $\mu$ is not a constant of motion only for the
   P SSC.
   
   When one ignores the spin of the test-body, and confines oneself to the 
   geodesic approximation, then for the Kerr background there is another conserved
   quantity \cite{Carter:1968c} called Carter constant. This constant appears to
   be a unique characteristic of the Kerr spacetime \cite{Markakis:2014h}. 
   For the MP a Carter-like constant has been found only in the linear-in-spin
   approximation of the T SSC \cite{Rudiger:1981c,Rudiger:1983c}. But, for a
   linear in spin Hamiltonian formulation introduced in \cite{Barausse:2009xi}
   this appears not to be the case \cite{Kunst:2016tla}. Thus, the conservation
   of the Carter-like quantities is an open question for the MP equations.
   
   \subsection{ Circular equatorial orbits (CEOs)} \label{sec:CEO}
   
   In Sec.~2~B of~Paper~\RM{1}, we have already given a detailed description
   of the procedures that we use to find CEOs and ISCOs for a spinning particle for the three SSCs.
   The formulas were stated in a form valid for any SAR spacetime
   so they can also be applied to the present Kerr case and are therefore omitted here.
   \footnote{In appendix~\ref{sec:OKScor} we provide two formulas that were
   presented in \cite{Harms:2016d} just for the Kerr spacetime in the respective
   SAR generalized versions.}
   
   We only recall that, for a given pair of $(r, S)$,
   a spinning particle CEO is uniquely characterized by the four quantities
   $~v^t,~v^\phi,~p^t,~p^\phi$. Equivalently, either $v^t$ or $v^\phi$ can be replaced
   by the orbital frequency $\Omega=d\phi / dt = v^\phi/v^t$. 
   We have computed these quantities for a set of $(r,S)$ and for $\ha=\pm0.9$.
   The results are presented in
   Tabs.~\ref{Tab:comparison_EOM_Omega&vut_a0p9}-\ref{Tab:comparison_EOM_put&puphi_am0p9}.
   Looking at these tables, we see that at large orbital radii $r$ the
   discrepancies between the SSCs are small to non-existent, at least for the number
   of digits given here. As expected, the discrepancies grow when we approach
   the central Kerr BH and the curvature increases, because the differences
   between the described test-particles, that are entailed by different choices
   of the SSC, become relevant through the spin-curvature coupling. We note that,
   at a given radius, the discrepancies are smaller for $\ha=0.9$ than for
   $\ha=-0.9$. This can be connected to the results on the shift of the ISCO to
   smaller radii for $\ha>0$, i.e.~$r_{\rm ISCO}(\ha=+0.9) < r_{\rm ISCO}(\ha=-0.9)$,
   see next section and cf.~Tabs.~\ref{tab:ISCOs_xa0p9}-~\ref{tab:ISCOs_xa0m9}.

   \section{ISCO} \label{sec:ISCO}
 
 \begin{figure}[t]
  \centering  
  {\includegraphics[width=0.45\textwidth]{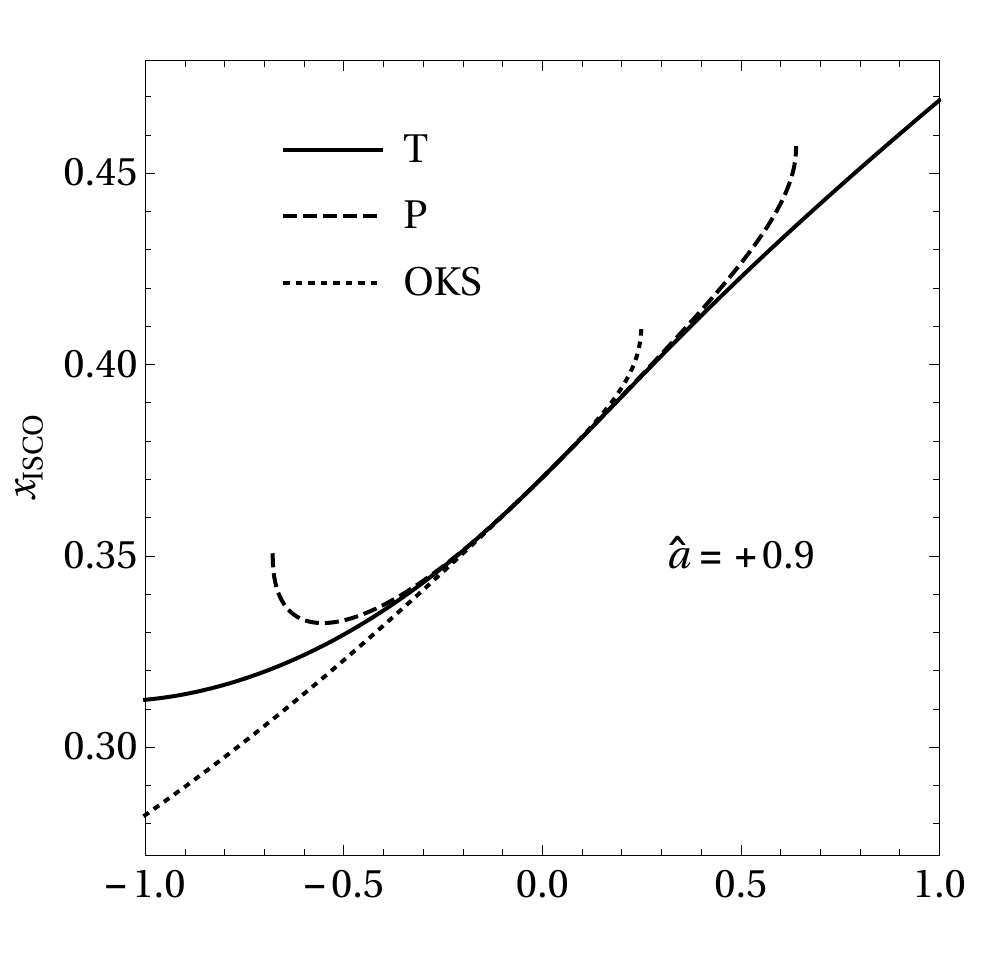} 
  \includegraphics[width=0.45\textwidth]{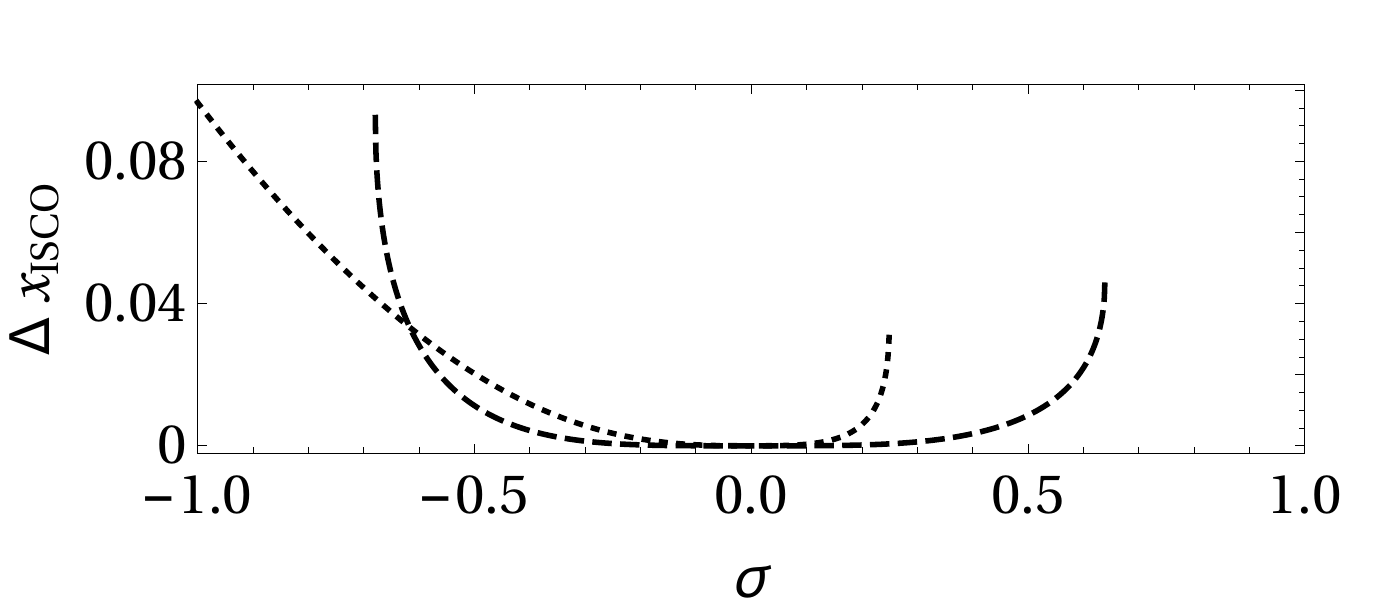} }
  \\
  \caption{Top panel: The frequency parameter at the ISCO radius,
    $x_\textrm{ISCO}=x(r_\textrm{ISCO})$, as a function of the
    particle's spin $\sigma$ for the Kerr spin parameter $\ha=0.9$.
    Bottom panel: Relative differences $ \Delta x_\textrm{ISCO} $
    of the ISCO's frequency parameter with respect to the one of the T SSC.    
    For the P SSC (long dashes) the ISCO computations fail for spins $\sigma <-0.68$
    and $\sigma >0.64$, for the OKS SSC (short dashes) the ISCO computations fail
    for spins $\sigma >0.25$.  
  } 
  \label{fig:xSa0p9}
\end{figure}

\begin{figure}[t]
  \centering  
  {\includegraphics[width=0.45\textwidth]{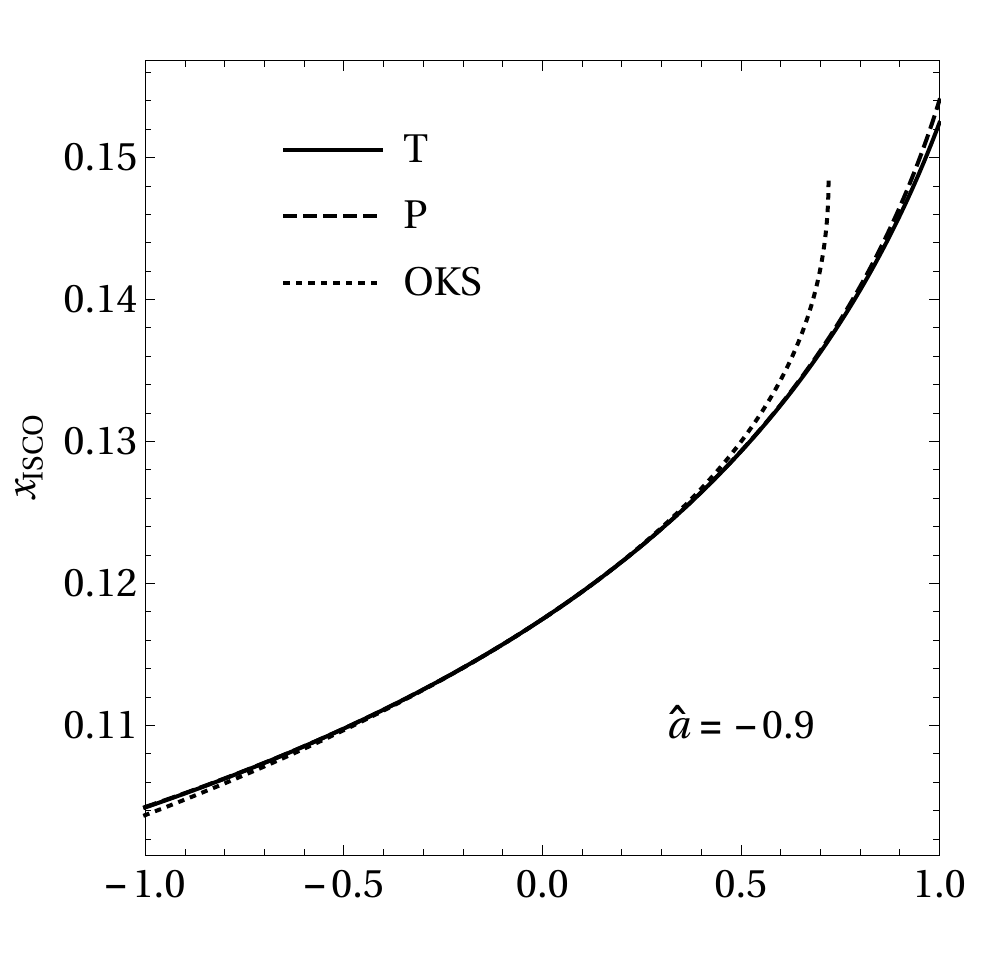} 
  \includegraphics[width=0.45\textwidth]{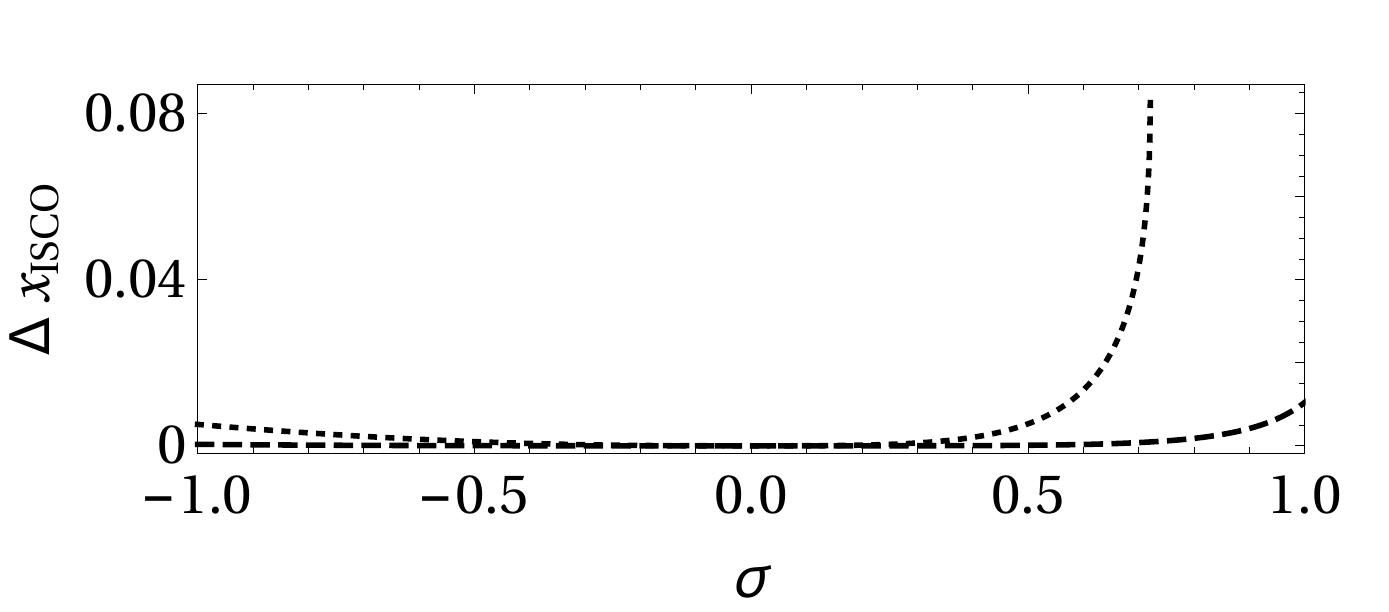} }
  \\
  \caption{Top panel: The ISCO frequency parameter $x_\textrm{ISCO}$ as a
    function of the particle's spin $\sigma$ for the Kerr spin parameter
    $\ha=-0.9$. 
    Bottom panel: Relative differences $  \Delta x_\textrm{ISCO} $
    of the ISCO's frequency parameter with respect to the one of the T SSC.    
    For the OKS SSC (short dashes) the ISCO computations fail for
    spins $\sigma >0.721$.  
  } 
  \label{fig:xSam0p9}
\end{figure}

\begin{table}[t]
\caption{Frequency parameter $x_\textrm{ISCO}$ at the ISCO of a spinning
  particle on a Kerr background with $\hat{a}=0.9$ computed for different
  SSCs. Entries with backslash $/$ mean that the
  ISCO values for these configurations could not be found, see text.}
\centering
\begin{ruledtabular}
  \begin{tabular}[t]{c c c c c } 
  $\sigma$   &
  $x_{\rm ISCO}^{\rm T }$  &
  $x_{\rm ISCO}^{\rm P }$  &  
  $x_{\rm ISCO}^{\rm OKS }$  \\
  \hline
  \hline
  $ 0.90  $  &  0.460219  &  $/$       &  $/$   \\ 
  $ 0.70  $  &  0.442008  &  $/$       &  $/$    \\ 
  $ 0.50  $  &  0.422794  &  0.426418  &  $/$   \\ 
  $ 0.30  $  &  0.402267  &  0.402715  &  $/$   \\ 
  $ 0.10  $  &  0.380886  &  0.380892  &  0.381057  \\ 
  $ -0.10  $ &  0.360472  &  0.360478  &  0.360370  \\ 
  $ -0.30  $ &  0.343009  &  0.343460  &  0.341071  \\ 
  $ -0.50  $ &  0.329397  &  0.333122  &  0.322706  \\ 
  $ -0.70  $ &  0.319706  &  $/$       &  0.305499  \\ 
  $ -0.90  $ &  0.313832  &  $/$       &  0.289632  \\ 
  \end{tabular} 
 \end{ruledtabular}
\label{tab:ISCOs_xa0p9}
\end{table}

\begin{table}[t]
\caption{Frequency parameter $x_\textrm{ISCO}$ at the ISCO of a spinning particle
  on a Kerr background with $\hat{a}=-0.9$ computed for different SSCs.
  Entries with backslash $/$ mean that the  ISCO values for these configurations
  could not be found, see text.}
\centering
\begin{ruledtabular}
  \begin{tabular}[t]{c c c c c } 
  $\sigma$   &
  $x_{\rm ISCO}^{\rm T }$  &
  $x_{\rm ISCO}^{\rm P }$  &  
  $x_{\rm ISCO}^{\rm OKS }$  \\
  \hline
  \hline
  $ 0.90  $  &  0.145882  &  0.146490  &  $/$   \\ 
  $ 0.70  $  &  0.136267  &  0.136373  &  0.142137  \\ 
  $ 0.50  $  &  0.129279  &  0.129295  &  0.129963  \\ 
  $ 0.30  $  &  0.123831  &  0.123833  &  0.123916  \\ 
  $ 0.10  $  &  0.119400  &  0.119400  &  0.119402  \\ 
  $ -0.10  $ &  0.115690  &  0.115690  &  0.115689  \\ 
  $ -0.30  $ &  0.112519  &  0.112519  &  0.112488  \\ 
  $ -0.50  $ &  0.109765  &  0.109768  &  0.109654  \\ 
  $ -0.70  $ &  0.107344  &  0.107355  &  0.107099  \\ 
  $ -0.90  $ &  0.105195  &  0.105221  &  0.104767  \\ 
  \end{tabular} 
 \end{ruledtabular}
\label{tab:ISCOs_xa0m9}
\end{table}
       
   The position of the ISCO is of importance, since it provides a notion of the 
   regime where most of the orbits are stable. Moreover, the ISCO is a gauge
   independent notion in the sense that an orbit cannot be stable for one
   observer, but unstable for another one. Moreover, to provide a fully gauge
   invariant discussion one can dump the orbital radius as the central parameter
   by arguing in terms of the orbital frequency parameter as measured by an
   observer at infinity. We therefore prefer to work with 
   \begin{align} \label{eq:freqPar}
     x\equiv (M~\Omega)^{2/3}
   \end{align}
   where $\Omega$ is the orbital frequency. Note though that the spin $\sigma$
   remains in a sense gauge dependent (SSC dependent), as argued at the end of
  this section.
      
   For the comparison we have chosen relatively large values of the Kerr
   parameter, $\ha=\pm 0.9$, in order to make the impact on the position
   of the ISCO prominent. To argue quantitatively, we also compute the relative differences
   \begin{align}\label{eq:RelDif}
     \Delta x_\textrm{ISCO} = (x_{\rm{ISCO}}^{\rm{SSC}} - x_{\rm{ISCO}}^{\rm{T}})/x_{\rm{ISCO}}^{\rm{T}}
   \end{align}
   of the ISCO frequency parameter of a SSC with respect to the ISCO frequency
   parameter of the T SSC. 
   
   Both Figs.~\ref{fig:xSa0p9},~\ref{fig:xSam0p9} show  how the three SSCs
   converge to the same ISCO frequency as the geodesic limit
   $\sigma \rightarrow 0$ is approached. Actually, the bottom panel of
   Fig.~\ref{fig:xSa0p9} shows that for $\ha=0.9$, the relative differences are
   $\Delta x_\textrm{ISCO}<0.5\%$ in the range $|\sigma|<0.2$ 
   for both the P and the OKS SSC, while the bottom panel of Fig.~\ref{fig:xSam0p9}
   shows that for $\ha=-0.9$, the relative differences are below
   $\Delta x_\textrm{ISCO}<0.5\%$ for all spins up to $\sigma<0.4$. For $\ha=0.9$
   Fig.~\ref{fig:xSa0p9} illustrates how drastically the SSCs diverge as large
   values of $|\sigma|$ are reached. For the P SSC (long dashes) our numerical
   procedures could only find ISCOs in the spin range $-0.68<\sigma<0.62$, while
   for the OKS SSC (short dashes) the ISCO computations fail for spins
   $\sigma >0.25$. The abrupt changes in the inclinations of the P and OKS SSC
   curves indicate that either there are no ISCOs outside the stated ranges of
   $\sigma$, or that we reach a cusp after which our computational
   approach fails to find ISCOs. This failure was also present for the
   Schwarzschild case, see discussion Sec.~\RM{4}~B in Paper~\RM{1}.
   On the other hand, for $\ha=-0.9$ all the discrepancies are mild as long
   as $\sigma<0.4$ (Fig.~\ref{fig:xSam0p9}) and only for the OKS SSC 
   (short dashes) there appears to be no ISCO when $\sigma >0.721$.
   It is worthy mentioning that only for the T SSC we could find
   ISCOs for all the range of $\sigma$.
    
   One could take the view that the discrepancies we see in 
   Figs.~\ref{fig:xSa0p9},~\ref{fig:xSam0p9} come from the fact that 
   the different dynamics are not describing the same physical body, since
   the transformation laws for translating from one centroid to another
   centroid~\cite{Kyrian:2007zz,Costa:?} have not been applied.
   However, one should keep in mind that if these transformation laws were
   applied then from a circular orbit we  would likely obtain a non-circular
   orbit, since not only the spin $\sigma$ would change but also the position of
   the centroid. Thus, if for one centroid the worldline is on the ISCO,
   for the other centroid it is not. In order to understand what is going on,
   one has to recall that we are trying to describe an extended body in its
   pole-dipole approximation, i.e.~only by its mass and spin. In general, however,
   an extended body has an infinite number of multipoles that are consciously
   neglected in our approach. Since even the quadrupole terms are neglected, 
   the body is implicitly assumed to be free of tidal deformations. Thus, two
   points of the same physical body lying at different radial distances cannot
   even tidally react to the gradients of the gravitational field.
   This is the reason why the SSCs can be interpreted as a gauge transformation 
   for the pole-dipole approximation in a flat spacetime, but in a curved 
   spacetime, if one starts from two centroids describing the same physical body,
   after a while the worldlines can get outside the body's worldtube
   \cite{Kyrian:2007zz,Costa:?}. In a few words, the discrepancies between the
   SSCs in a curved spacetime result from the fact that we try to describe an
   extended body by its two first multipoles.

 \section{Asymptotic Gravitational Wave fluxes} \label{sec:GWFlux}

\begin{figure*}[t]
  \centering  
  \includegraphics[width=0.45\textwidth]{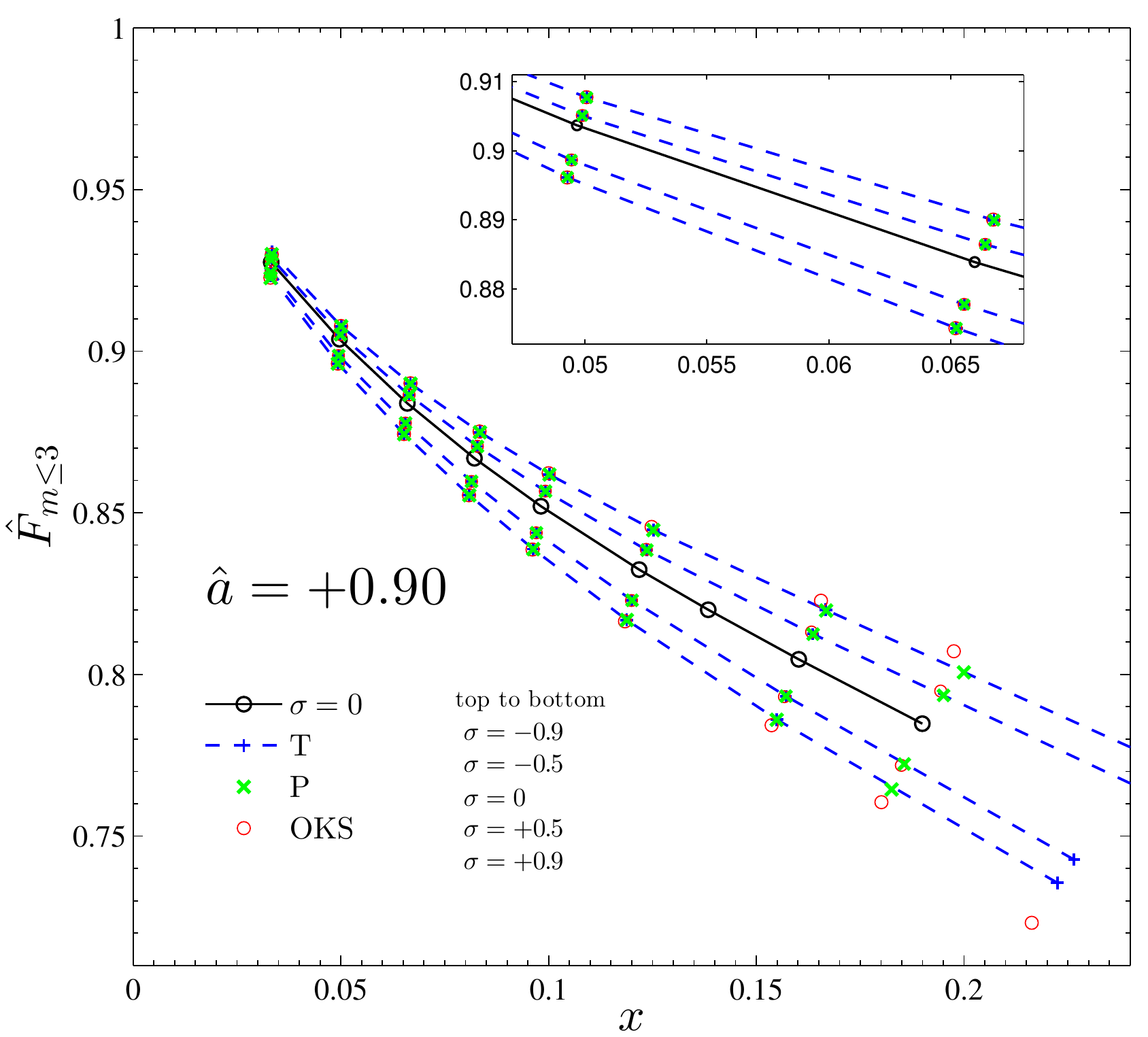} 
  \includegraphics[width=0.45\textwidth]{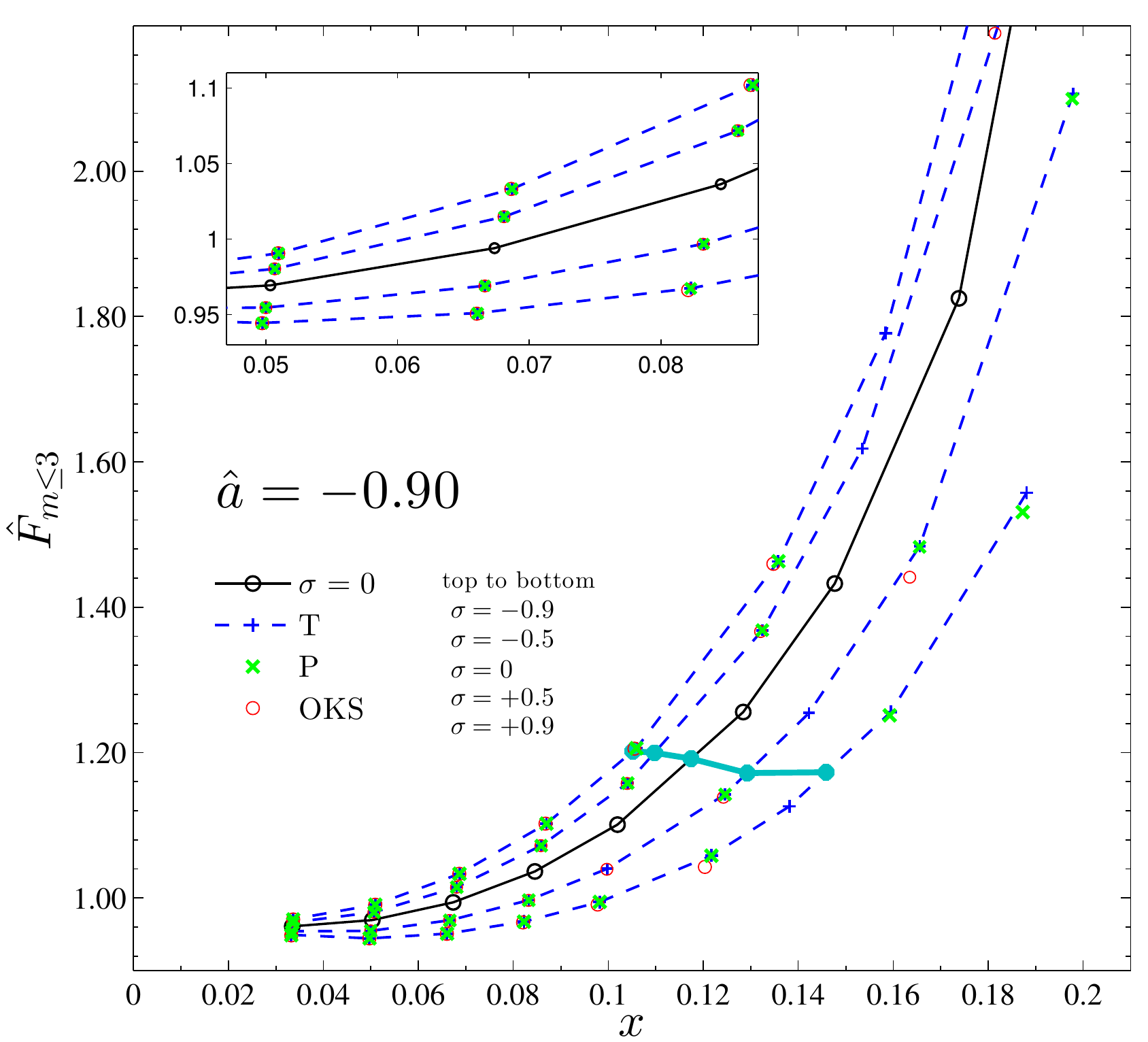} 
  \\
  \caption{
   Comparison of the GW energy flux, approximated by the sum of the $m=1,2,3$
   modes, containing all $\ell$ contributions,  over the frequency parameter $x$.
   We compare the fluxes of three different SSCs from the circular dynamics of a
   particle with spins $\sigma=0,\pm 0.5,\pm 0.9$ around a Kerr black hole
   with spin $\hat{a}=0.9$ (left panel) and spin $\hat{a}=-0.9$ (right panel). The different
   cases in the plots are illustrated as follows: blue dashed lines with pluses for the T SSC,
   green crosses for the P SSC, red circles for the OKS SSC, and the solid black line
   with circles for the nonspinning particle limit. The fluxes at the ISCOs
   are connected along the different spins $\sigma$ for the T case (thick cyan, not in plotted range on left panel).
   }
  \label{fig:VisualFluxComparison}
\end{figure*}

\begin{figure*}[t]
  \centering  
  \includegraphics[width=0.45\textwidth]{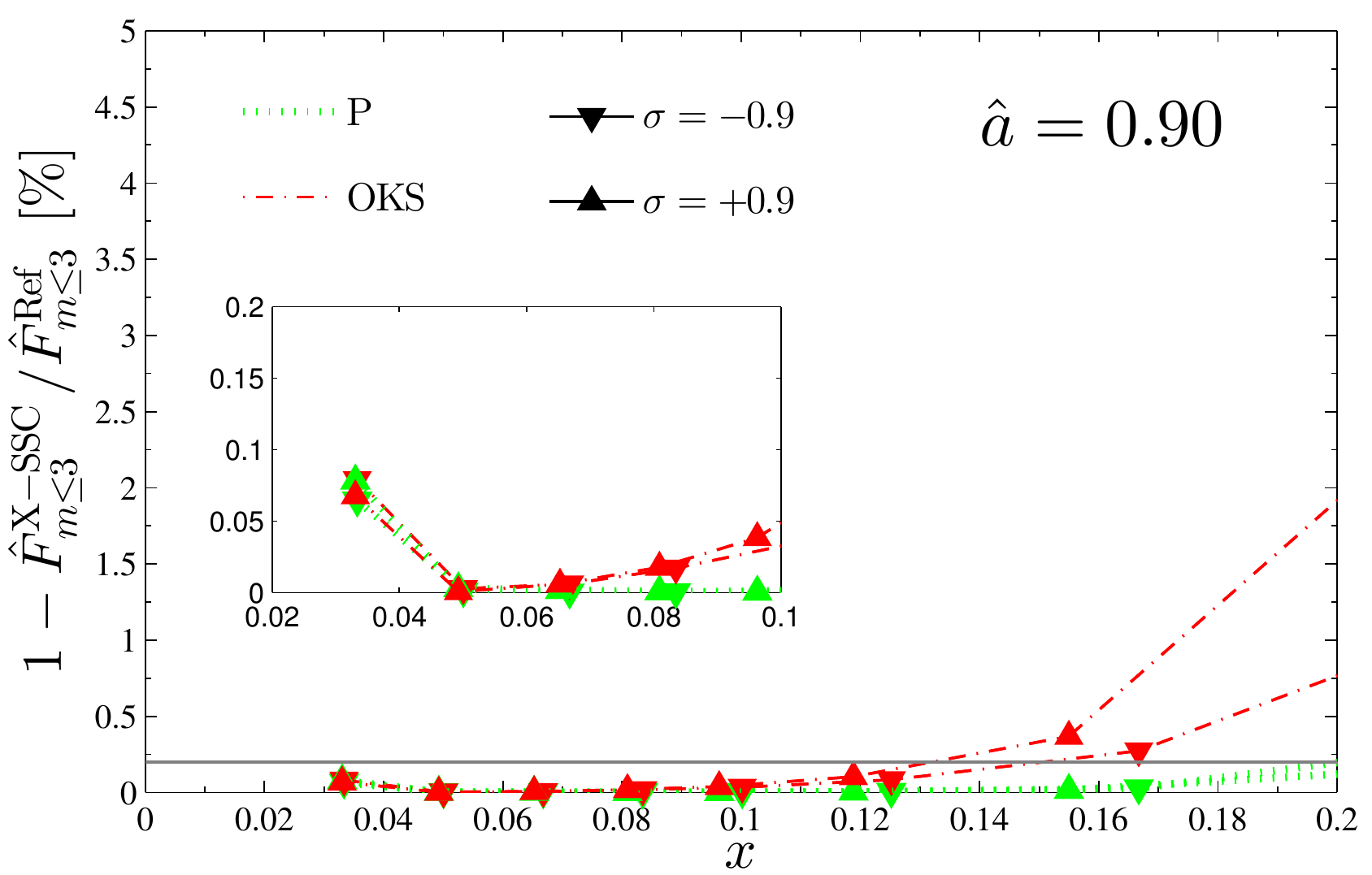} 
  \includegraphics[width=0.45\textwidth]{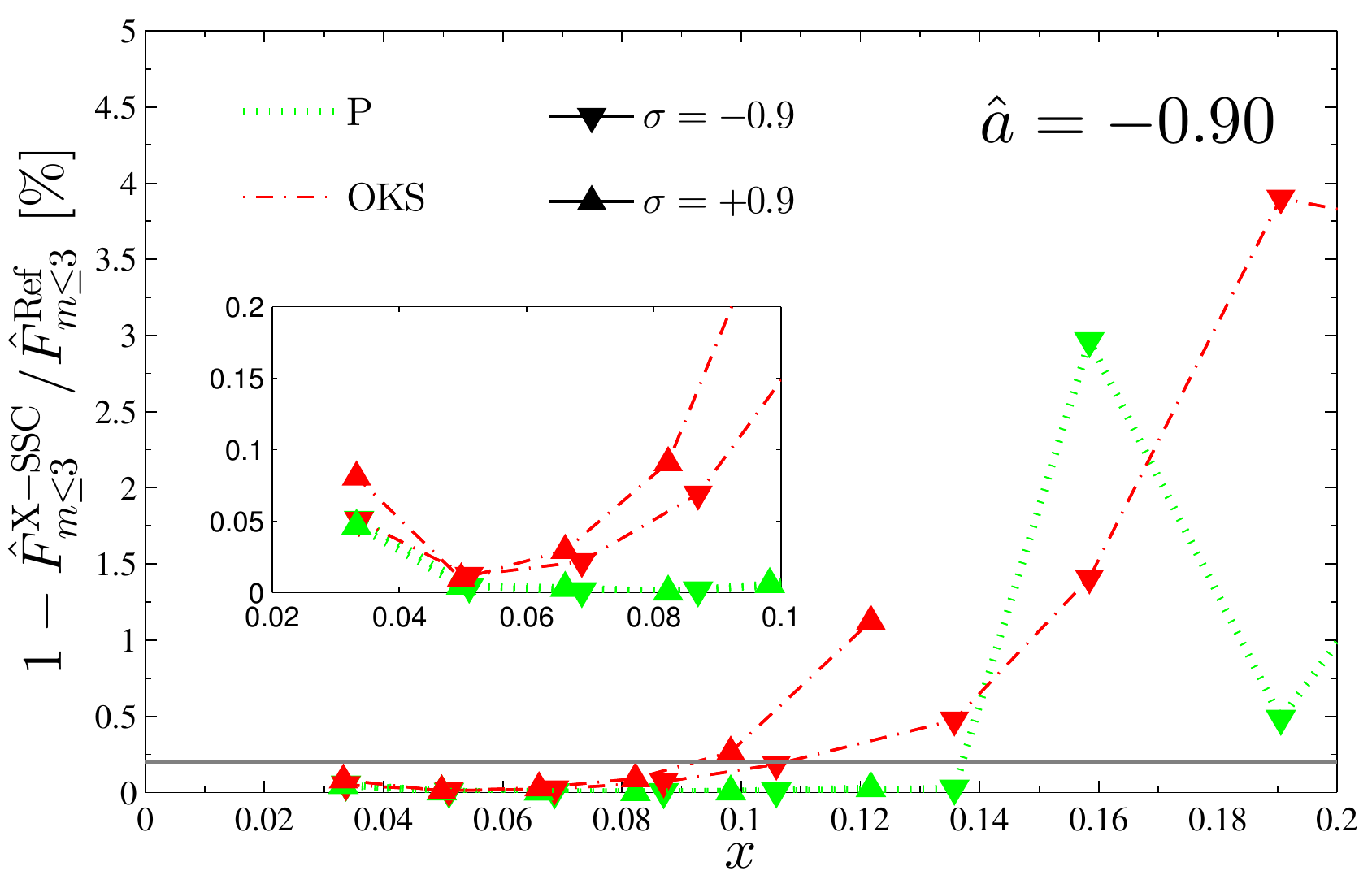} 
  \\
  \caption{
  Relative differences in the full GW flux, approximated as the sum over the $m=1,2,3$ modes,
  using the P SSC (green, dotted), and the OKS SSC (red, dash-dotted)
  with respect to the T SSC reference case.
  The left panel presents the $\hat{a}=0.9$ case, while the right the 
  $\hat{a}=-0.9$ case.
  We consider the two dimensionless particle spins $\sigma=-0.9$ (downward triangles) and $\sigma=0.9$ (upward triangles).   
  The gray horizontal lines at $0.2 \%$ mark our estimated relative numerical accuracy.  
   }
  \label{fig:QuantitativeFluxComparison}
\end{figure*}

 Let us now analyze the differences implicated by different SSCs
 from another perspective, namely by comparing the 
 fluxes of energy that are emitted to infinity in the form of gravitational waves
 as the spinning-particle moves along the CEOs.
 
 The waves have been calculated by feeding the dynamical
 quantities of~Tabs.~\ref{Tab:comparison_EOM_Omega&vut_a0p9}-\ref{Tab:comparison_EOM_put&puphi_am0p9}
 to the \texttt{Teukode}.
 This code solves the Teukolsky equation (TE)~\cite{Teukolsky:1972my,Teukolsky:1973ha}
 for a given particle perturbation of the Kerr spacetime in the time-domain,
 using a (2+1)D form and hyperboloidal slices~\cite{Calabrese:2005rs,Zenginoglu:2007jw,Zenginoglu:2009hd,Zenginoglu:2010cq,Vano-Vinuales:2014koa}.
 More details on the numerical methods employed in the \texttt{Teukode}
 can be found in~\cite{Harms:2013ib,Harms:2014dqa,Nagar:2014kha,Harms:2015ixa},
 as well as comparisons against independent literature results
 \cite{Bernuzzi:2012ku,Bernuzzi:2010xj,
Barausse:2011kb,Sundararajan:2010sr,Sundararajan:2007jg,Hughes:1999bq,Taracchini:2014zpa}.
 The particular numerical setup, including resolution and convergence rates,
 that was used for the experiments of this work is
 described in Sec.\RM{5}~A of Paper~\RM{1}. We only recall that we estimate
 our flux results to have a relative numerical accuracy of $~0.2\%$. 
 We have considered CEOs at BL-radii in the set
 $\hat{r} \in \{4,5,6,8,10,12,20,30\}$
 and for four particle spins~$\sigma=\pm0.5,\pm0.9$.

 It is convenient to decompose the GW fluxes in terms of a spin-weighted spherical harmonic basis 
 \begin{align}
   \label{eq:Flux}
  F &= \sum_{m=1}^{\infty} F_{m} 
   = \sum_{\ell=2}^{\infty}\sum_{m=1}^{m=\ell} F_{\ell m} \nonumber \quad ,
%    = \dfrac{2}{16\pi}\sum_{\ell=2}^{\infty}\sum_{m=1}^{m=\ell}(m\Omega)^2| r h_{\ell m}|^2 \quad,
 \end{align}
%  where we follow the notation of~\cite{Damour:2008gu}. 
 where $F_m$ and $F_{\ell m}$ are defined to contain both the $m$ and $-m$ contributions
 (both are equivalent for GWs from a particle on a CEO). Our waveform algorithm
 directly provides the fluxes $F_m$ with all $\ell$-contributions included.
 Following standard practice in the discussion of fluxes, we normalize them according to
 \be
  \hat{F}_{m} = \begin{cases}
  \frac{F_m}{F^{\rm LO}_{2m}} & m=1 \quad, \vspace{4 pt}\\ 
  \frac{F_m}{F^{\rm LO}_{mm}} & \textrm{otherwise} \quad ,
 \end{cases} 
 \ee
 using the leading-order (LO) flux $ F^{\rm LO}_{\ell m}(x) $ given by the
 quadrupole formula as a normalization factor. Thus the fluxes should approach $1$
 as $r\to \infty$ or $x\to 0$.

 We have computed the fluxes for the three dominant multipoles ${m=1,2,3}$,
 having in mind to provide a reasonable approximation of the full flux.
 The results for $\hat{F}_{m\le 3}$ are plotted as functions of the
 frequency parameter $x$ for each SSC in Fig.~\ref{fig:VisualFluxComparison}.
 To back a quantitative comparison, the corresponding
 relative differences of the P and OKS SSC fluxes with respect to the T SSC
 flux are shown in Fig.~\ref{fig:QuantitativeFluxComparison}. 
 The data used for the plots is given in Tab.~\ref{Tab:comparison_fluxes_Fm123_apm9}.
 
 The main observation conveyed in Fig.~\ref{fig:VisualFluxComparison} is that
 the discrepancies in the fluxes between the three SSCs 
 are small over the whole considered parameter range
 and tend to vanish at large orbital distances, i.e.~at small frequencies $x\to 0$.
 This was the central conclusion made in Paper~\RM{1} for the Schwarzschild case
 and we can now prove that it holds for the Kerr case as well.
 While the panels in Fig.~\ref{fig:VisualFluxComparison} only visually indicate
 a remarkable agreement of all fluxes, the relative differences
 shown in the panels in Fig.~\ref{fig:QuantitativeFluxComparison} prove that at small $x$ the
 different SSCs in fact result in equivalent test-particle descriptions with respect to the gravitational fluxes,
 at least within our numerical accuracy limits of $0.2 \%$ (horizontal gray line).
 The differences remain below this limit even up to radii as small as $r\gtrsim 8M$ for $\ha=0.9$ and $r\gtrsim 10M$ for $\ha=-0.9$.
 In fact, looking closer at the left panel of Fig.~\ref{fig:QuantitativeFluxComparison},
 the discrepancies between the T and the P SSC (green dotted lines)
 are below the estimated numerical accuracy over the whole
 range of the frequency parameter for $\hat{a}=0.9$. 
 The discrepancies between the T and the OKS SSC grow beyond the numerical
 accuracy threshold for $x>0.14$. On the other hand, the right panel of 
 Fig.~\ref{fig:QuantitativeFluxComparison} shows that for $\ha=-0.9$ this threshold is reached 
 already at $x \approx 0.1$  for the relative differences between the T and the OKS SSC (red dash-dotted),
 and at $x \approx 0.14$ for the relative differences between the T and the P SSC (green dotted).
 Overall, we conclude that the T and the P SSC are almost indistinguishable for most of the considered cases.
 As expected, the differences grow as the central object is approached.
 We note that at a given frequency these discrepancies are smaller for the $\hat{a}=0.9$ case
%  (left panels of Figs.~\ref{fig:VisualFluxComparison}-\ref{fig:QuantitativeFluxComparison})
 than for the $\hat{a}=-0.9$ case.
%  (right panels of Figs.~\ref{fig:VisualFluxComparison}-\ref{fig:QuantitativeFluxComparison}).
 These observations are compliant with our analysis of the orbital parameters as discussed earlier
 (Tabs.~\ref{Tab:comparison_EOM_Omega&vut_a0p9}-\ref{Tab:comparison_EOM_put&puphi_am0p9}).

 \section{Conclusions} \label{sec:Concl}
 
 In this article circular equatorial orbits of a spinning test-body on a Kerr
 background have been studied numerically and their asymptotic gravitational wave fluxes 
 have been computed for three SSCs, i.e.
 i)~MP with the T SSC,
 ii)~MP with the P SSC,
 iii)~MP with the OKS SSC.
 For the P SSC and the OKS SSC this is the first time that CEOs and ISCOs, as
 well as gravitational fluxes, have been calculated on a Kerr background. 
 
 Summarizing our analyses in just one sentence, we have found that the influence
 of the SSC on the orbital dynamics as well as the fluxes is negligible for
 small orbital frequency parameters, i.e.~for large orbital distances.
 Significant discrepancies arise when the orbital frequencies are big. 
 Thus, the central conclusion made in Paper~\RM{1} for the Schwarzschild
 background is proved true for the Kerr background. Moreover, at a given
 frequency these discrepancies for CEO parameters and fluxes tend to be milder
 for positive Kerr parameters than for negative Kerr parameters. Contrarily,
 when looking at ISCO shifts, the picture is inversed, i.e. for $\hat{a}=-0.9$
 the discrepancies between the different descriptions are generally smaller than
 for $\ha=+0.9$. We assume that this behavior can be explained by the fact that
 ISCOs for $\ha=+0.9$ lie at smaller radial distances ($r<3.5$) than for $\ha=-0.9$
 ($r>6$), i.e.~ISCOs for $\ha=+0.9$ lie in a gravitationally stronger regime
 than for $\ha=-0.9$. 
%  This is probably also the reason of the picture inversion, since the CEOs and 
%  the fluxes we studied have smaller frequency parameters ($x<0.28$) than the
%  ISCOs for $\ha=+0.9$.
 
 \begin{table*}
 
\caption{ Comparison of the dynamical quantities $\Omega$ and $v^t$
          obtained within the MP formalism for 
          circular, equatorial orbits of a spinning particle
          around a Kerr BH with $\hat{a}=0.9$ for three different SSCs:
          i)~T SSC,
          ii)~P SSC,
          iii)~OKS SSC.
          The different cases are indicated 
          as subscripts in the respective quantities.
          The values are normalized by setting
          $\mu=M=1$.
          Note that $v^\ph$ can be computed from the given
          quantities.
        }
\label{Tab:comparison_EOM_Omega&vut_a0p9}

  \begin{tabular}[t]{| c | c | c c c | c c c |}
 \hline 
 \multicolumn{8}{|l|}{ {\bf\large{$\hat{a}=0.9$}} } \\ \hline 
 $\hat{r}$ & 
 $\sigma $ & 
 $ M \Omega_{\rm{T}}  $  &  $ M \Omega_{\rm{P}}  $  &  $ M \Omega_{\rm{OKS}}  $  & 
 $ v_{\rm{T}}^t        $  &  $ v_{\rm{P}}^t        $  &  $ v_{\rm{OKS}}^t        $    
 \\ 
 \hline 
4.00   & -0.90 & 0.12401 & 0.12475  & 0.12053  & 1.69791  & 1.70412  & 1.67043 \\ 
       & -0.50 & 0.11825 & 0.11834  & 0.11716  & 1.65359  & 1.65428  & 1.64587 \\ 
       & 0.50  & 0.10778 & 0.10772  & 0.10656  & 1.58703  & 1.58672  & 1.58032 \\ 
       & 0.90  & 0.10493 & 0.10466  & 0.10058  & 1.57157  & 1.57015  & 1.54974 \\ 
  \hline 
5.00   & -0.90 & 0.08924 & 0.08941  & 0.08779  & 1.48065  & 1.48179  & 1.47158 \\ 
       & -0.50 & 0.08613 & 0.08616  & 0.08568  & 1.46155  & 1.46170  & 1.45889 \\ 
       & 0.50  & 0.07997 & 0.07996  & 0.07949  & 1.42793  & 1.42785  & 1.42551 \\ 
       & 0.90  & 0.07809 & 0.07800  & 0.07643  & 1.41866  & 1.41823  & 1.41090 \\ 
  \hline 
6.00   & -0.90 & 0.06807 & 0.06812  & 0.06738  & 1.36626  & 1.36659  & 1.36239 \\ 
       & -0.50 & 0.06620 & 0.06621  & 0.06598  & 1.35595  & 1.35600  & 1.35480 \\ 
       & 0.50  & 0.06229 & 0.06229  & 0.06207  & 1.33612  & 1.33609  & 1.33503 \\ 
       & 0.90  & 0.06101 & 0.06098  & 0.06025  & 1.33009  & 1.32993  & 1.32662 \\ 
  \hline 
8.00   & -0.90 & 0.04430 & 0.04431  & 0.04409  & 1.24699  & 1.24776  & 1.24663 \\ 
       & -0.50 & 0.04347 & 0.04350  & 0.04340  & 1.24349  & 1.24349  & 1.24316 \\ 
       & 0.50  & 0.04162 & 0.04162  & 0.04155  & 1.23455  & 1.23455  & 1.23424 \\ 
       & 0.90  & 0.04096 & 0.04096  & 0.04074  & 1.23154  & 1.23151  & 1.23053 \\  
  \hline 
10.00  & -0.90 & 0.03171 & 0.03171  & 0.03163  & 1.18689  & 1.18689  & 1.18647 \\ 
       & -0.50 & 0.03127 & 0.03127  & 0.03125  & 1.18468  & 1.18468  & 1.18455 \\ 
       & 0.50  & 0.03026 & 0.03026  & 0.03023  & 1.17977  & 1.17977  & 1.17965 \\ 
       & 0.90  & 0.02988 & 0.02988  & 0.02980  & 1.17803  & 1.17802  & 1.17764 \\  
  \hline 
12.00  & -0.90 & 0.02412 & 0.02412  & 0.02409  & 1.14994  & 1.14995  & 1.14975 \\ 
       & -0.50 & 0.02386 & 0.02386  & 0.02385  & 1.14862  & 1.14862  & 1.14856 \\ 
       & 0.50  & 0.02325 & 0.02325  & 0.02323  & 1.14558  & 1.14558  & 1.14553 \\ 
       & 0.90  & 0.02301 & 0.02301  & 0.02298  & 1.14448  & 1.14447  & 1.14429 \\ 
  \hline 
15.00  & -0.90 & 0.01726 & 0.01726  & 0.01724  & 1.11558  & 1.11558  & 1.11550 \\ 
       & -0.50 & 0.01712 & 0.01712  & 0.01711  & 1.11485  & 1.11485  & 1.11483 \\ 
       & 0.50  & 0.01679 & 0.01679  & 0.01678  & 1.11316  & 1.11316  & 1.11313 \\ 
       & 0.90  & 0.01666 & 0.01665  & 0.01665  & 1.11252  & 1.11252  & 1.11245 \\ 
  \hline 
20.00  & -0.90 & 0.01120 & 0.01120  & 0.01120  & 1.08357  & 1.08357  & 1.08355 \\ 
       & -0.50 & 0.01114 & 0.01114  & 0.01114  & 1.08323  & 1.08323  & 1.08323 \\ 
       & 0.50  & 0.01100 & 0.01100  & 0.01100  & 1.08242  & 1.08242  & 1.08242 \\ 
       & 0.90  & 0.01094 & 0.01094  & 0.01094  & 1.08211  & 1.08211  & 1.08209 \\  
  \hline 
30.00  & -0.90 & 0.00609 & 0.00609  & 0.00609  & 1.05374  & 1.05374  & 1.05373 \\ 
       & -0.50 & 0.00608 & 0.00608  & 0.00608  & 1.05362  & 1.05362  & 1.05362 \\ 
       & 0.50  & 0.00603 & 0.00603  & 0.00603  & 1.05333  & 1.05333  & 1.05333 \\ 
       & 0.90  & 0.00601 & 0.00601  & 0.00601  & 1.05322  & 1.05322  & 1.05321 \\ 
  \hline 
 \end{tabular} 

\end{table*}

\begin{table*}
 
 \caption{Comparison of the dynamical quantities $p^t$ and $\hat{p}^\phi$  for 
          circular, equatorial orbits of a spinning particle
          around a Kerr BH with $\hat{a}=0.9$. 
          See caption of Tab~\ref{Tab:comparison_EOM_Omega&vut_a0p9}
          for details.
          }
 \label{Tab:comparison_EOM_put&puphi_a0p9}
  
 \begin{tabular}[t]{| c | c | c c c  | c c c |}
 \hline 
 \multicolumn{8}{|l|}{ {\bf\large{$\hat{a}=0.9$}} } \\ \hline 
 $\hat{r}$ & 
 $\sigma $ & 
 $ p_{\rm{T}}^t $  &  $ p_{\rm{P}}^t $ &  $ p_{\rm{OKS}}^t $ & 
   $ \hat{p}_{\rm{T}}^\phi $  &  $\hat{p}_{\rm{P}}^\phi $ & $\hat{p}_{\rm{OKS}}^\phi $   
 \\ 
 \hline 
4.00  & -0.90 & 1.67837 & 1.67954  & 1.67043  &  0.20403 &  0.20455 & 0.20134 \\ 
      & -0.50 & 1.64857 & 1.64877  & 1.64587  &  0.19377 &  0.19385 & 0.19282 \\ 
      & 0.50  & 1.58369 & 1.58353  & 1.58032  &  0.16974 &  0.16967 & 0.16840 \\ 
      & 0.90  & 1.56221 & 1.56136  & 1.54974  &  0.16110 &  0.16078 & 0.15587 \\  
  \hline 
5.00  & -0.90 & 1.47439 & 1.47466  & 1.47158  &  0.13011 &  0.13022 & 0.12920 \\ 
      & -0.50 & 1.45983 & 1.45988  & 1.45889  &  0.12531 &  0.12533 & 0.12500 \\
      & 0.50  & 1.42664 & 1.42659  & 1.42551  &  0.11372 &  0.11371 & 0.11331 \\ 
      & 0.90  & 1.41487 & 1.41464  & 1.41090  &  0.10936 &  0.10927 & 0.10784 \\ 
  \hline 
6.00  & -0.90 & 1.36363 & 1.36372  & 1.36239  &  0.09218 &  0.09221 & 0.09179 \\ 
      & -0.50 & 1.35521 & 1.35522  & 1.35480  &  0.08953 &  0.08953 & 0.08939 \\ 
      & 0.50  & 1.33551 & 1.33550  & 1.33503  &  0.08302 &  0.08302 & 0.08286 \\ 
      & 0.90  & 1.32827 & 1.32819  & 1.32662  &  0.08051 &  0.08049 & 0.07993 \\ 
  \hline 
8.00  & -0.90 & 1.24699 & 1.24700  & 1.24663  &  0.05507 &  0.05507 & 0.05496 \\ 
      & -0.50 & 1.24328 & 1.24328  & 1.24316  &  0.05399 &  0.05399 & 0.05396 \\ 
      & 0.50  & 1.23437 & 1.23436  & 1.23424  &  0.05132 &  0.05132 & 0.05128 \\ 
      & 0.90  & 1.23097 & 1.23096  & 1.23053  &  0.05027 &  0.05027 & 0.05014 \\ 
  \hline 
10.00 & -0.90 & 1.18661 & 1.18662  & 1.18647  &  0.03757 &  0.03757 & 0.03753 \\ 
      & -0.50 & 1.18460 & 1.18460  & 1.18455  &  0.03702 &  0.03702 & 0.03701 \\ 
      & 0.50  & 1.17970 & 1.17970  & 1.17965  &  0.03567 &  0.03567 & 0.03566 \\ 
      & 0.90  & 1.17780 & 1.17780  & 1.17764  &  0.03514 &  0.03514 & 0.03509 \\ 
  \hline 
12.00 & -0.90 & 1.14982 & 1.14982  & 1.14975  &  0.02771 &  0.02771 & 0.02409 \\ 
      & -0.50 & 1.14858 & 1.14858  & 1.14856  &  0.02740 &  0.02740 & 0.02739 \\ 
      & 0.50  & 1.14555 & 1.14555  & 1.14553  &  0.02662 &  0.02662 & 0.02662 \\ 
      & 0.90  & 1.14436 & 1.14436  & 1.14429  &  0.02631 &  0.02631 & 0.02629 \\ 
  \hline 
15.00 & -0.90 & 1.11553 & 1.11553  & 1.11550  &  0.01924 &  0.01924 & 0.01923 \\ 
      & -0.50 & 1.11484 & 1.11484  & 1.11483  &  0.01908 &  0.01908 & 0.01908 \\ 
      & 0.50  & 1.11314 & 1.11314  & 1.11313  &  0.01868 &  0.01868 & 0.01868 \\ 
      & 0.90  & 1.11248 & 1.11248  & 1.11245  &  0.01853 &  0.01853 & 0.01852 \\ 
  \hline 
20.00 & -0.90 & 1.08356 & 1.08356  & 1.08355  &  0.01214 &  0.01214 & 0.01214 \\ 
      & -0.50 & 1.08323 & 1.08323  & 1.08323  &  0.01207 &  0.01207 & 0.01207 \\ 
      & 0.50  & 1.08242 & 1.08242  & 1.08242  &  0.01190 &  0.01190 & 0.01190 \\ 
      & 0.90  & 1.08210 & 1.08210  & 1.08209  &  0.01183 &  0.01183 & 0.01183 \\ 
  \hline 
30.00 & -0.90 & 1.05374 & 1.05374  & 1.05373  &  0.00642 &  0.00642 & 0.00642 \\ 
      & -0.50 & 1.05362 & 1.05362  & 1.05362  &  0.00640 &  0.00640 & 0.00640 \\ 
      & 0.50  & 1.05333 & 1.05333  & 1.05333  &  0.00635 &  0.00635 & 0.00635 \\ 
      & 0.90  & 1.05321 & 1.05321  & 1.05321  &  0.00633 &  0.00633 & 0.00633 \\ 
  \hline 
 \end{tabular}

\end{table*}
 
\begin{table*}
 
\caption{ Comparison of the dynamical quantities $\Omega$ and $v^t$ for 
          circular, equatorial orbits of a spinning particle
          around a Kerr BH with $\hat{a}=-0.9$. 
          See caption of Tab~\ref{Tab:comparison_EOM_Omega&vut_a0p9}
          for details.
        }
\label{Tab:comparison_EOM_Omega&vut_am0p9}

  \begin{tabular}[t]{| c | c | c c c | c c c |}
 \hline 
 \multicolumn{8}{|l|}{ {\bf\large{$\hat{a}=-0.9$}} } \\ \hline 
 $\hat{r}$ & 
 $\sigma $ & 
 $ M \Omega_{\rm{T}}  $  &  $ M \Omega_{\rm{P}}  $  &  $ M \Omega_{\rm{OKS}}  $  & 
 $ v_{\rm{T}}^t        $  &  $ v_{\rm{P}}^t        $  &  $ v_{\rm{OKS}}^t        $    
 \\ 
  \hline 
5.00   & -0.90 & 0.11771 & 0.12235  & 0.10993  & 2.55529  & 2.87875  & 2.20837 \\ 
       & -0.50 & 0.10798 & 0.10824  & 0.10526  & 2.14402  & 2.15238  & 2.06446 \\ 
       & 0.50  & 0.08802 & 0.08791  &    /     & 1.72996  & 1.72841  &    /    \\ 
       & 0.90  & 0.08157 & 0.08105  &    /     & 1.64988  & 1.64402  &    /    \\ 
  \hline 
6.00   & -0.90 & 0.08318 & 0.08350  & 0.08053  & 1.66615  & 1.67129  & 1.62679 \\ 
       & -0.50 & 0.07817 & 0.07821  & 0.07729  & 1.59478  & 1.59534  & 1.58355 \\ 
       & 0.50  & 0.06738 & 0.06735  & 0.06612  & 1.47781  & 1.47756  & 1.46661 \\ 
       & 0.90  & 0.06370 & 0.06356  &    /     & 1.44639  & 1.44525  &    /    \\ 
  \hline 
8.00   & -0.90 & 0.05004 & 0.05007  & 0.04947  & 1.33074  & 1.33100  & 1.32612 \\ 
       & -0.50 & 0.04819 & 0.04820  & 0.04801  & 1.31612  & 1.31615  & 1.31472 \\ 
       & 0.50  & 0.04400 & 0.04400  & 0.04379  & 1.28638  & 1.28635  & 1.28500 \\ 
       & 0.90  & 0.04249 & 0.04247  & 0.04175  & 1.27669  & 1.27656  & 1.27216 \\  
  \hline 
10.00  & -0.90 & 0.03447 & 0.03448  & 0.03429  & 1.22388  & 1.22392  & 1.22264 \\ 
       & -0.50 & 0.03360 & 0.03360  & 0.03354  & 1.21812  & 1.21812  & 1.21774 \\ 
       & 0.50  & 0.03155 & 0.03155  & 0.03149  & 1.20551  & 1.20551  & 1.20514 \\ 
       & 0.90  & 0.03079 & 0.03079  & 0.03058  & 1.20111  & 1.20108  & 1.19989 \\  
  \hline 
12.00  & -0.90 & 0.02565 & 0.02565  & 0.02558  & 1.17021  & 1.17022  & 1.16974 \\ 
       & -0.50 & 0.02517 & 0.02517  & 0.02515  & 1.16726  & 1.16726  & 1.16711 \\ 
       & 0.50  & 0.02403 & 0.02403  & 0.02400  & 1.16056  & 1.16056  & 1.16042 \\ 
       & 0.90  & 0.02360 & 0.02359  & 0.02351  & 1.15814  & 1.15813  & 1.15767 \\ 
  \hline 
15.00  & -0.90 & 0.01801 & 0.01801  & 0.01798  & 1.12573  & 1.12573  & 1.12557 \\ 
       & -0.50 & 0.01777 & 0.01777  & 0.01776  & 1.12434  & 1.12434  & 1.12429 \\ 
       & 0.50  & 0.01721 & 0.01721  & 0.01720  & 1.12108  & 1.12108  & 1.12103 \\ 
       & 0.90  & 0.01699 & 0.01699  & 0.01696  & 1.11986  & 1.11986  & 1.11971 \\ 
  \hline 
20.00  & -0.90 & 0.01150 & 0.01150  & 0.01145  & 1.08794  & 1.08794  & 1.08790 \\ 
       & -0.50 & 0.01141 & 0.01141  & 0.01141  & 1.08737  & 1.08737  & 1.08736 \\ 
       & 0.50  & 0.01118 & 0.01118  & 0.01118  & 1.08601  & 1.08601  & 1.08600 \\ 
       & 0.90  & 0.01109 & 0.01109  & 0.01108  & 1.08546  & 1.08546  & 1.08545 \\  
  \hline 
30.00  & -0.90 & 0.00618 & 0.00618  & 0.00618  & 1.05515  & 1.05515  & 1.05514 \\ 
       & -0.50 & 0.00615 & 0.00615  & 0.00615  & 1.05497  & 1.05497  & 1.05497 \\ 
       & 0.50  & 0.00609 & 0.00609  & 0.00609  & 1.05455  & 1.05455  & 1.05454 \\ 
       & 0.90  & 0.00606 & 0.00606  & 0.00606  & 1.05438  & 1.05438  & 1.05437 \\ 
  \hline 
 \end{tabular} 

\end{table*}

\begin{table*}
 
 \caption{Comparison of the dynamical quantities $p^t$ and $\hat{p}^\phi$ for 
          circular, equatorial orbits of a spinning particle
          around a Kerr BH with $\hat{a}=-0.9$. 
          See caption of Tab~\ref{Tab:comparison_EOM_Omega&vut_a0p9}
          for details.
          }
 \label{Tab:comparison_EOM_put&puphi_am0p9}
  
 \begin{tabular}[t]{| c | c | c c c  | c c c |}
 \hline 
 \multicolumn{8}{|l|}{ {\bf\large{$\hat{a}=-0.9$}} } \\ \hline 
 $\hat{r}$ & 
 $\sigma $ & 
 $ p_{\rm{T}}^t $  &  $ p_{\rm{P}}^t $ &  $ p_{\rm{OKS}}^t $ & 
   $ \hat{p}_{\rm{T}}^\phi $  &  $ \hat{p}_{\rm{P}}^\phi $ &  $ \hat{p}_{\rm{OKS}}^\phi    $   
 \\ 
  \hline 
5.00  & -0.90 & 2.38518 & 2.48830  & 2.20837  & 0.27281  & 0.29121  & 0.24276 \\ 
      & -0.50 & 2.11446 & 2.11928  & 2.06446  & 0.22628  & 0.22716  & 0.21731 \\ 
      & 0.50  & 1.71885 & 1.71752  &     /    & 0.14989  & 0.14961  &    /    \\ 
      & 0.90  & 1.62324 & 1.61784  &     /    & 0.12840  & 0.12721  &    /    \\ 
  \hline 
6.00  & -0.90 & 1.64602 & 1.64816  & 1.62679  & 0.13474  & 0.13519  & 0.13101 \\ 
      & -0.50 & 1.58977 & 1.59006  & 1.58355  & 0.12365  & 0.12372  & 0.12239 \\ 
      & 0.50  & 1.47471 & 1.47452  & 1.46661  & 0.09886  & 0.09882  & 0.09697 \\ 
      & 0.90  & 1.43797 & 1.43705  &     /    & 0.09007  & 0.08986  &    /    \\ 
  \hline 
8.00  & -0.90 & 1.32813 & 1.32823  & 1.32612  & 0.06603  & 0.06606  & 0.06560 \\ 
      & -0.50 & 1.31538 & 1.31540  & 1.31472  & 0.06327  & 0.06327  & 0.06312 \\ 
      & 0.50  & 1.28579 & 1.28578  & 1.28500  & 0.05646  & 0.05646  & 0.05627 \\ 
      & 0.90  & 1.27496 & 1.27488  & 1.27216  & 0.05381  & 0.05379  & 0.05311 \\ 
  \hline 
10.00 & -0.90 & 1.22315 & 1.22316  & 1.22264  & 0.04203  & 0.04203  & 0.04192 \\ 
      & -0.50 & 1.21790 & 1.21791  & 1.21774  & 0.04088  & 0.04088  & 0.04084 \\ 
      & 0.50  & 1.20533 & 1.20532  & 1.20514  & 0.03799  & 0.03799  & 0.03795 \\ 
      & 0.90  & 1.20054 & 1.20052  & 1.19989  & 0.03685  & 0.03685  & 0.03669 \\ 
  \hline 
12.00 & -0.90 & 1.16992 & 1.16993  & 1.16974  & 0.02996  & 0.02996  & 0.02992 \\ 
      & -0.50 & 1.16717 & 1.16718  & 1.16711  & 0.02936  & 0.02936  & 0.02935 \\ 
      & 0.50  & 1.16049 & 1.16049  & 1.16042  & 0.02787  & 0.02787  & 0.02785 \\ 
      & 0.90  & 1.15790 & 1.15789  & 1.15767  & 0.02727  & 0.02727  & 0.02722 \\ 
  \hline 
15.00 & -0.90 & 1.12563 & 1.12563  & 1.12557  & 0.02025  & 0.02025  & 0.02024 \\ 
      & -0.50 & 1.12431 & 1.12431  & 1.12429  & 0.01997  & 0.01997  & 0.01997 \\ 
      & 0.50  & 1.12105 & 1.12105  & 1.12103  & 0.01928  & 0.01928  & 0.01928 \\ 
      & 0.90  & 1.11978 & 1.11978  & 1.11971  & 0.01901  & 0.01901  & 0.01899 \\ 
  \hline 
20.00 & -0.90 & 1.08791 & 1.08791  & 1.08790  & 0.01251  & 0.01251  & 0.01251 \\ 
      & -0.50 & 1.08736 & 1.08736  & 1.08736  & 0.01240  & 0.01240  & 0.01240 \\ 
      & 0.50  & 1.08600 & 1.08600  & 1.08600  & 0.01214  & 0.01214  & 0.01214 \\ 
      & 0.90  & 1.08546 & 1.08546  & 1.08545  & 0.01203  & 0.01203  & 0.01203 \\ 
  \hline 
30.00 & -0.90 & 1.05514 & 1.05514  & 1.05514  & 0.00652  & 0.00652  & 0.00652 \\ 
      & -0.50 & 1.05497 & 1.05497  & 1.05497  & 0.00649  & 0.00649  & 0.00649 \\ 
      & 0.50  & 1.05454 & 1.05454  & 1.05454  & 0.00642  & 0.00642  & 0.00642 \\ 
      & 0.90  & 1.05437 & 1.05437  & 1.05437  & 0.00639  & 0.00639  & 0.00639 \\ 
  \hline 
 \end{tabular}

\end{table*}

\begin{table*} 
\caption{
        Comparison of the full energy fluxes, approximated as the sum
        over the $m=1,2,3$ modes containing all $\ell$-contributions,
        produced by a spinning particle in circular motion around a spinning black-hole
        with $\ha=-0.9$ (left table) and $\ha=+0.9$ (right table).
        The values for the energy fluxes have to be understood as normalized by
        the leading order Newtonian flux.
        The table compares the fluxes
        at several Boyer-Lindquist radii $r$,
        for the four particle spins $\sigma=\pm0.9\pm0.5$,
        and for three different circular dynamics: 
        i)~MP with the T SSC,
        ii)~MP with the P SSC,
        iii)~MP with the OKS SSC.
        We use the Tulczyjew case as the reference when computing
        the respective differences shown in the $\Delta[\%]$ columns.
        In case the relative differences fall below the level of $0.001 \%$
        we do just write $<0.001 \%$ to avoid citing more digits.          
        If a certain combination was not simulated we write a backslash $/$.
        The T SSC results for $r=30M$ were obtained at higher resolutions
        than all the other cases, see discussion in Sec.\RM{5}A of Paper~\RM{1},
        which is why the relative differences are not consistent 
        and thus shown in brackets.
        The main observation is that the relative differences between the respective fluxes
        vanish as the orbital distance grows.
        At $r=20M$ the energy fluxes from all dynamics agree
        in all measured cases up to $\lesssim 0.01 \%$ or better.
        }
\label{Tab:comparison_fluxes_Fm123_apm9}
   
\begin{tabular}[t]{| c | c | c c c c c  |}
 \hline 
 \multicolumn{7}{|l|}{ {\bf\large{$\hat{a}=-0.90$}} } \\ \hline 
 $\hat{r}$ & 
 $\sigma $ & 
 $\hat{F}_{m \leq 3}^{\rm{T}}$   &  $\hat{F}_{m \leq 3}^{\rm{P}}$   & $ \Delta[\%] $ & 
 $\hat{F}_{m \leq 3}^{\rm{OKS}}$ & $ \Delta[\%] $ \\ 
  \hline 
4.00  & -0.90  & /  & /  & /  & /  & /  \\ 
  & -0.50  & /  & /  & /  & /  & /  \\ 
  & 0.50  & /  & /  & /  & /  & /  \\ 
  & 0.90  & /  & /  & /  & /  & /  \\ 
  \hline 
5.00  & -0.90  & 6.478  & 7.059  & 8.968  & 5.467  & 15.614  \\ 
  & -0.50  & 4.661  & 4.690  & 0.615  & 4.372  & 6.208  \\ 
  & 0.50  & 2.107  & 2.100  & 0.335  & /  & /  \\ 
  & 0.90  & 1.558  & 1.531  & 1.704  & /  & /  \\ 
  \hline 
6.00  & -0.90  & 2.569  & 2.577  & 0.297  & 2.501  & 2.674  \\ 
  & -0.50  & 2.215  & 2.217  & 0.056  & 2.190  & 1.151  \\ 
  & 0.50  & 1.484  & 1.483  & 0.061  & 1.441  & 2.878  \\ 
% & 0.90  & 1.256  & 1.251  & 0.361  & 1.011  & 19.507  \\  % data not trustworthy for OKS, a=-0.9, sig=+0.9, r<=6
  & 0.90  & 1.256  & 1.251  & 0.361  & /  & /  \\ 
  \hline 
8.00  & -0.90  & 1.463  & 1.463  & 0.011  & 1.460  & 0.230  \\ 
  & -0.50  & 1.368  & 1.368  & 0.003  & 1.367  & 0.128  \\ 
  & 0.50  & 1.142  & 1.142  & 0.006  & 1.139  & 0.333  \\ 
  & 0.90  & 1.058  & 1.058  & 0.039  & 1.043  & 1.491  \\ 
  \hline 
10.00  & -0.90  & 1.206  & 1.206  & $ < $ 0.001  & 1.205  & 0.032  \\ 
  & -0.50  & 1.158  & 1.158  & $ < $ 0.001  & 1.158  & 0.026  \\ 
  & 0.50  & 1.040  & /  & /  & 1.039  & 0.080  \\ 
  & 0.90  & 0.994  & 0.994  & 0.008  & 0.991  & 0.351  \\ 
  \hline 
12.00  & -0.90  & 1.102  & 1.102  & $ < $ 0.001  & 1.102  & 0.001  \\ 
  & -0.50  & 1.072  & 1.072  & $ < $ 0.001  & 1.072  & 0.007  \\ 
  & 0.50  & 0.997  & 0.997  & $ < $ 0.001  & 0.997  & 0.028  \\ 
  & 0.90  & 0.967  & 0.967  & 0.001  & 0.966  & 0.117  \\ 
  \hline 
15.00  & -0.90  & 1.033  & 1.033  & $ < $ 0.001  & 1.033  & 0.002  \\ 
  & -0.50  & 1.015  & 1.015  & $ < $ 0.001  & 1.015  & $ < $ 0.001  \\ 
  & 0.50  & 0.969  & 0.969  & $ < $ 0.001  & 0.969  & 0.007  \\ 
  & 0.90  & 0.951  & 0.951  & 0.003  & 0.951  & 0.035  \\ 
  \hline 
20.00  & -0.90  & 0.991  & 0.991  & 0.004  & 0.991  & 0.008  \\ 
  & -0.50  & 0.980  & 0.980  & 0.005  & 0.980  & 0.007  \\ 
  & 0.50  & 0.955  & 0.955  & 0.005  & 0.955  & 0.003  \\ 
  & 0.90  & 0.944  & 0.944  & 0.005  & 0.944  & 0.010  \\ 
  \hline 
30.00  & -0.90  & 0.971  & 0.970  & (0.051)  & 0.970  & (0.051)  \\ 
  & -0.50  & 0.966  & 0.965  & (0.070)  & 0.965  & (0.068)  \\ 
  & 0.50  & 0.954  & 0.953  & (0.059)  & 0.953  & (0.066)  \\ 
  & 0.90  & 0.949  & 0.949  & (0.047)  & 0.948  & (0.081)  \\ 
  \hline 
 \end{tabular} 
 \begin{tabular}[t]{| c | c | c c c c c  |}
 \hline 
 \multicolumn{7}{|l|}{ {\bf\large{$\hat{a}=0.90$}} } \\ \hline 
 $\hat{r}$ & 
 $\sigma $ & 
 $\hat{F}_{m \leq 3}^{\rm{T}}$   &  $\hat{F}_{m \leq 3}^{\rm{P}}$   & $ \Delta[\%] $ & 
 $\hat{F}_{m \leq 3}^{\rm{OKS}}$ & $ \Delta[\%] $ \\ 
 \hline 
4.00  & -0.90  & 0.772  & /  & /  & 0.786  & 1.755  \\ 
  & -0.50  & 0.766  & /  & /  & /  & /  \\ 
  & 0.50  & 0.743  & /  & /  & /  & /  \\ 
  & 0.90  & 0.736  & /  & /  & 0.723  & 1.683  \\ 
  \hline 
5.00  & -0.90  & /  & 0.801  & /  & 0.807  & /  \\ 
  & -0.50  & /  & 0.794  & /  & 0.795  & /  \\ 
  & 0.50  & /  & 0.772  & /  & 0.772  & /  \\ 
  & 0.90  & /  & 0.764  & /  & 0.760  & /  \\ 
  \hline 
6.00  & -0.90  & 0.820  & 0.820  & 0.034  & 0.823  & 0.344  \\ 
  & -0.50  & 0.813  & 0.812  & 0.003  & 0.813  & 0.075  \\ 
  & 0.50  & 0.793  & 0.793  & $ < $ 0.001  & 0.793  & 0.020  \\ 
  & 0.90  & 0.786  & 0.786  & 0.008  & 0.784  & 0.223  \\ 
  \hline 
8.00  & -0.90  & 0.845  & 0.845  & 0.006  & 0.846  & 0.112  \\ 
  & -0.50  & 0.838  & 0.838  & $ < $ 0.001  & 0.839  & 0.025  \\ 
  & 0.50  & 0.823  & 0.823  & $ < $ 0.001  & 0.823  & 0.003  \\ 
  & 0.90  & 0.817  & 0.817  & 0.002  & 0.816  & 0.052  \\ 
  \hline 
10.00  & -0.90  & 0.862  & 0.862  & 0.002  & 0.862  & 0.047  \\ 
  & -0.50  & 0.857  & 0.857  & $ < $ 0.001  & 0.857  & 0.011  \\ 
  & 0.50  & 0.844  & 0.844  & $ < $ 0.001  & 0.844  & $ < $ 0.001  \\ 
  & 0.90  & 0.839  & 0.839  & $ < $ 0.001  & 0.839  & 0.015  \\ 
  \hline 
12.00  & -0.90  & 0.875  & 0.875  & $ < $ 0.001  & 0.875  & 0.025  \\ 
  & -0.50  & 0.871  & 0.871  & $ < $ 0.001  & 0.871  & 0.006  \\ 
  & 0.50  & 0.860  & 0.860  & $ < $ 0.001  & 0.860  & $ < $ 0.001  \\ 
  & 0.90  & 0.855  & 0.855  & $ < $ 0.001  & 0.855  & 0.005  \\ 
  \hline 
15.00  & -0.90  & 0.890  & 0.890  & $ < $ 0.001  & 0.890  & 0.010  \\ 
  & -0.50  & 0.886  & 0.886  & $ < $ 0.001  & 0.886  & 0.002  \\ 
  & 0.50  & 0.878  & 0.878  & 0.001  & 0.878  & $ < $ 0.001  \\ 
  & 0.90  & 0.874  & 0.874  & 0.001  & 0.874  & $ < $ 0.001  \\ 
  \hline 
20.00  & -0.90  & 0.908  & 0.908  & $ < $ 0.001  & 0.908  & 0.004  \\ 
  & -0.50  & 0.905  & 0.905  & $ < $ 0.001  & 0.905  & 0.002  \\ 
  & 0.50  & 0.899  & 0.899  & 0.003  & 0.899  & 0.003  \\ 
  & 0.90  & 0.896  & 0.896  & 0.003  & 0.896  & 0.001  \\ 
  \hline 
30.00  & -0.90  & 0.931  & 0.930  & (0.064)  & 0.930  & (0.079)  \\ 
  & -0.50  & 0.929  & 0.928  & (0.060)  & 0.928  & (0.054)  \\ 
  & 0.50  & 0.925  & 0.924  & (0.067)  & 0.924  & (0.086)  \\ 
  & 0.90  & 0.923  & 0.923  & (0.078)  & 0.923  & (0.067)  \\ 
  \hline 
 \end{tabular}
\end{table*}

\begin{acknowledgments}
G.L-G is supported by Grant No. GACR-17-06962Y.
Computations were partly performed on CINECA/Marconi under PRACE allocation
3522 (14th call, proposal number 2016153522). 

\end{acknowledgments}

\appendix

 \section{Formula corrections for CEOs under OKS SSC} \label{sec:OKScor}

  In Sec.~II.B.3 of \cite{Harms:2016d} eq.~(49) and the formula between 
  eqs.~(49)--(50) hold only for the Kerr spacetime, for a SAR spacetime on the 
  equatorial plane ($\theta=\pi/2$) they should read 
  \begin{align}
    \frac{d S^\theta}{d \lambda}+\frac{S^\theta v^r }{2 g_{\theta\theta}}\frac{\partial g_{\theta\theta}}{\partial r}=0 \quad, \nonumber \\
    \frac{d V^\theta}{d \lambda}+\frac{V^\theta v^r}{2 g_{\theta\theta}}\frac{\partial g_{\theta\theta}}{\partial r}=0 \quad, \nonumber
  \end{align}
  respectively.

%  \clearpage

\bibliographystyle{unsrt}
\bibliography{../../refs20170724/refs20170724}

\end{document}